%% file: preprint.tex
\documentclass[pre,final,notitlepage,lengthcheck,showpacs,superscriptaddress]{revtex4}
\begin{document}
\input epsf
\input{figures/figdef.tex}

\title{Random spread on the family of small-world networks }
\author{Sagar A. Pandit}
\email{sagar@prl.ernet.in}
\affiliation{Physical Research Laboratory, Ahmedabad 380 009, India}
\author{R. E. Amritkar}
\email{amritkar@prl.ernet.in}
\affiliation{Physical Research Laboratory, Ahmedabad 380 009, India}
\begin{abstract}
  We present the analytical and numerical results of a random walk on
  the family of small-world graphs. The average access time shows a
  crossover from the regular to random behavior with increasing
  distance from the starting point of the random walk. We introduce an 
  {\em independent step approximation}, which enables us to obtain
  analytic results for the average access time. We observe a scaling
  relation for the average access time in the degree of the nodes. The
  behavior of average access time as a function of $p$, shows
  striking similarity with that of the {\em characteristic length} of
  the graph. This observation may have important applications in
  routing and switching in networks with large number of nodes.  
\end{abstract}
\maketitle
\section{Introduction}

The small-world network exhibits unusual connection properties. On one
hand it shows strong clustering like regular graphs and on the other
hand it shows very small average shortest path between any two
nodes like random graphs. Watts and Strogatz have proposed a simple
model to describe 
small-world networks~\cite{WattsStrogatz}. The model gives a
prescription to generate a one parameter family of graphs, ranging
from the highly clustered (regular) graphs to the random graphs. 

Various properties of this model have been
studied~\cite{Sag,NewmanWatts,Moukarzel,MooreNewman}. The spread and
percolation properties investigated
in Refs.~\cite{Sag,NewmanWatts,Moukarzel,MooreNewman}, deal with the spread
of information (disease) along the shortest path in the graph or the
spread along the spanning tree.

In this paper, we study random walk on the family of small-world
networks. Such a random walk corresponds to {\em random spread} of
information on the network. In any realistic
application of the spread on a graph, we expect the spread to be
somewhere in between the two extremes viz. shortest path and the
random walk. e.g. In Milgram's experiment~\cite{Milgram}, which
studies the connection properties of social networks, the path of a
letter from a randomly chosen point to a fixed target is traced. The
only condition imposed on the transfer of letter was that, the letter
should be given to the person whom the sender knows by first name.
The path followed by such a letter would have both random as well as
shortest path elements in it. Another example is the path of internet
protocol packet which follows a similar algorithm for forwarding the
packet~\cite{Stevens}. 

The determination of shortest path between two nodes typically
requires $O(n^3)$ operations~\cite{Aho}, where $n$ is the number of
nodes. This process becomes prohibitively expensive as $n$ becomes
very large.  Examples of such networks are social networks, telephone
networks, internet etc. Another problem with the determination of
shortest path is the incomplete knowledge of the network. Hence, it is 
clear that an alternate method of generating path (which need not be
shortest) becomes necessary in these networks. Our analysis of random
walk shows that average access time between two nodes goes as $O(n)$,
for small-world geometry. It is benificial to consider a network
with random routing or switching, particularly if it has small-world
properties. Thus the random routing emerges to be both, practical and
computationally cheaper method for large networks.

In Section~\ref{Main} we discuss analytical and numerical results for
the {\em average access time} of the random walk on one parameter family of
graph ranging from regular case to the random case. We introduce an
{\em independent step approximation} which allows us to get analytical
expressions for the average access time. We discuss these results in
Section~\ref{discussion}. It is found that the random walk results are
similar to that of the shortest path results. Thus from the nature of
the outcome of an experiment it may be difficult to conclude whether
the spread was random or along the shortest
path. An important consequence of this result can be in the routing
and switching in very big networks. Random routing is a
promising method, particularly if the network has small-world
properties. Section~\ref{summary} summarizes the results.

\section{Random walk on small-world graphs}
\label{Main}

The random walk on a graph is performed as follows:
We start with a fixed node (say $i$) and at each step make a jump to
a node connected to $i$ with uniform probability $1/d(i)$, where
$d(i)$ is the degree of the node $i$, thus performing a random
walk. Such a random walk gives a finite {\em Markov
 chain}~\cite{Lovasz}. One of the most important quantities of interest 
in a finite Markov chain is the {\em average access time}. Let $D_{i,j}$ be
the average access time,
defined as the first passage time to the node $j$ if the walk starts
from the node $i$. We denote $D_j = D_{0,j}$. 

We perform the random walks on the family of graphs 
generated by the algorithm given by Watts and
Strogatz~\cite{WattsStrogatz}. The prescription gives a one parameter
family of graphs which interpolates between the regular case and the
random case. We refer to this family as the family of small-world
networks or graphs. The regular graph (denoted by $p=0$) is a graph with $n$
vertices on a circle with each node connected to $2k$ nearest neighbors.
The parameter $k$ is suitably chosen to keep the graph sparse but
connected. The other elements of the family are obtained by random
rewiring of each edge in the  
graph with probability $p$. It is seen that the small-world behavior
is prominent around the parameter value $p=0.01$. i.e., when only
$1\%$ edges are rewired. The case $p=1$ corresponds to the random
case~\footnote{In the rewiring algorithm of Watts and Strogatz the
  random graph defined by $p=1$ is not truly random. However, as far as
  spread is concern the small-world network ($p=0.01$) itself is good
  enough approximation to the random network. Hence, $p=1$ case is
  adequate to describe the random behavior.}.

\subsection{The regular case $(p=0)$}
\label{peqzero}

\begin{figure}[htbp]
  \begin{center}
    \def \epsfsize#1#2{0.45#1}
    \figdraw{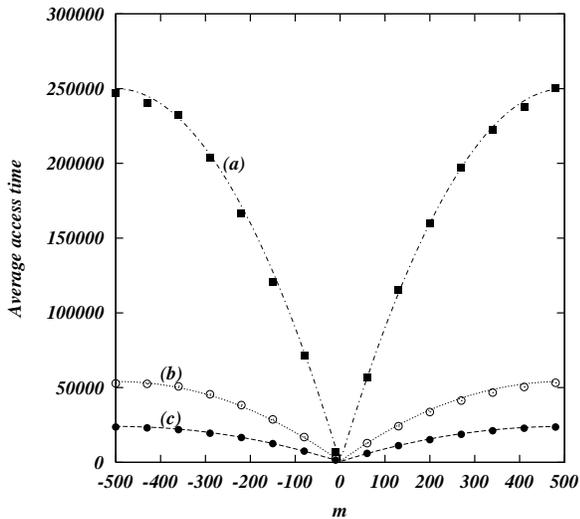}     
    \caption{The plot of average access time vs. the distance $m$ of a 
      site from the starting point ($m=0$)
    for the regular graphs. Curves (a), (b) and (c) correspond to
    $k=1$, $3$ and $5$ respectively. The points are the result of
    simulations of random walks on graph of size 1000. The lines are
    the analytical and scaling results obtained using
    equations~(\ref{eq:solkeqone}) and~(\ref{eq:block}).}
    \label{fig1}
  \end{center}
\end{figure}

Fig.~\ref{fig1} shows the results of the {\em average access time} in
simulation of random walk on the regular graph with $1000$ nodes for
several values of $k$. The average access time shows a linear behavior for
small $m$ and it shows quadratic nature for large $m$ due to the
circular topology. The lines are the analytic curves obtained as follows.

From the expectation values of conditional events, we can easily write
the recursion relation for average access time for the walk starting
from node $i$ as~\cite{Feller} 
\begin{eqnarray}
  \label{eq:rec}
  D_{i,m} = {1 \over {2k}} \sum_{j=1}^k ( D_{i+j,m} + D_{i-j,m} ) + 1
\end{eqnarray}
This is a $2k$-th order difference equation for $D_{i,m}$. We
first consider the case $k=1$. In this case the above recursion
relation reduces to a quadratic form given by 
\begin{eqnarray}
\label{eq:reckeqone}
  D_{i,m} = {1 \over 2} (D_{i+1,m} + D_{i-1,m}) + 1
\end{eqnarray}
The equation can be solved using standard techniques~\cite{Feller} and 
the solution is given by 
\begin{eqnarray}
  \label{eq:gensol}
  D_{i,m} = - (m - i)^2 + A\;(m-i) + B
\end{eqnarray}
where $A$ and $B$ are constants. The constants are determined using
the boundary conditions $D_{m,m} = 0$ and $D_{m-n,m} = 0$. Hence,
\begin{eqnarray}
  D_{i,m} = - i^2 + (2m -n) i - m^2 + mn \nonumber
\end{eqnarray}
Without loss of generality, we assume that the
 walk starts from $i=0$. So the {\em average access time} for site $m$
starting from zero is
\begin{eqnarray}
\label{eq:solkeqone}
  D_m = -m^2 + mn
\end{eqnarray}

Curve (a) in Fig.~\ref{fig1} shows both the analytical and numerical
results for the case $k=1$. The linear and the quadratic nature is
clearly seen. 

For general value of $k$, it is not possible to solve the difference
equation~(\ref{eq:rec}) exactly. The numerical simulations show that
the nature of the curves for different $k$ is the same as for
$k=1$. This suggests that there may be a scaling relation for a
general $k$. Using numerical data fit we find that the following
scaling relation fits the data reasonably well.
\begin{eqnarray}
  \label{eq:block}
  D_m^{k}(n) \approx (1 + \mu \ln(k)) D_{{m \over k}}^1({n \over k}) 
\end{eqnarray}
where $D_m^{k}$ is the average access time  from the site $0$ to site
$m$ on the graph with $2k$ nearest neighbors, $\mu = 0.86$~\footnote{ 
  We have also tried other types of scaling relations in particular
  scaling relation of the form $D_m^{k}(n) \approx k^\mu D_{{m \over
      k}}^1({n \over k})$. However, we find that the
  relation~(\ref{eq:block}) gives better fit to the numerical data.}.

A possible explanation for such a scaling can be a quotienting of a
graph by sub-graph of size $k$. 

\begin{figure}[htbp]
  \begin{center}
    \def \epsfsize#1#2{0.45#1}
    \figdraw{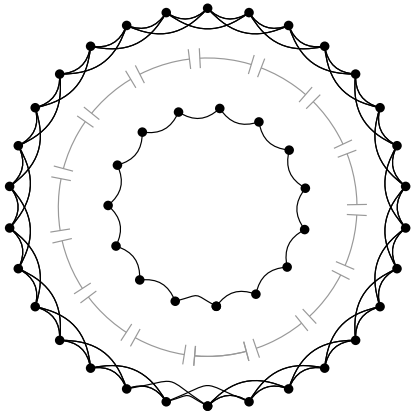}     
    \caption{Figure describes the quotienting procedure of reducing
      the regular graph with $n$ nodes and $2k$ neighbors to a graph
      with ${n \over k}$ nodes and $2$ neighbors. The outer
    graph is of 30 nodes and the inner quotiented graph is of 15 nodes.}
    \label{fig2}
  \end{center}
\end{figure}

The Fig~\ref{fig2}
explains the quotienting procedure clearly. The outer graph in
Fig~\ref{fig2} is of size $30$ and the inner graph is the quotiented
graph which has only nearest neighbor connections. While the walk on 
the quotiented graph is described by equation~(\ref{eq:solkeqone}),
the coefficient probably comes from the average time spent in each
block. Note that the number of blocks that the walker has to pass is
${m \over k}$.

From Fig.~\ref{fig1} we see that the scaling relation~(\ref{eq:block}) 
shows an excellent matching with the numerical results for various values of 
$k$.

\subsection{The random case $(p=1)$}
\label{peqone}

For the completely random case the access time becomes independent of
$m$ and $D_{i,j} = n-1$, $\forall$ $i$, $j$. This result can be
obtained as follows. 

We note that to calculate average access time one must consider an
ensemble of graph for a given $p$.
 The average access time is obtained by first averaging
over several realizations of random walk on a given graph and then it
is averaged over various members of the ensemble of graphs. We now introduce
an {\em independent step approximation} where we assume that the order 
of these two averages is interchanged. Thus in this approximation each
step of the random walk is averaged over all the realizations of the
graphs. This approximation is a
kind of mean field approximation done in statistical mechanics.

Let $P_{i,j}$ be the probability of reaching site $j$ from site $i$ in 
one step. It is obvious that $\sum_i P_{i,j} = 1$. Using independent
step approximation, we get
\begin{eqnarray}
  P_{i,j} = {1 \over {n-1}} \nonumber
\end{eqnarray}
Thus the probability of reaching a site $m$ at any time step is ${1
  \over {n-1}}$. Hence, the probability of reaching site $m$ for the
first time in $t$ time steps is
\begin{eqnarray}
  \label{eq:prob}
  P_m(t) = \left( 1 - {1 \over {n-1}} \right)^{t-1} {1 \over {n-1}}
\end{eqnarray}
Thus $D_m$ is given by
\begin{eqnarray}
  D_m &=& \sum_t t P_m(t) \nonumber \\
  &=& n-1
\end{eqnarray}

\subsection{The intermediate case $(0 < p < 1)$}

\begin{figure}[hbt]
  \begin{center}
 \def \epsfsize#1#2{0.45#1}
 \figdraw{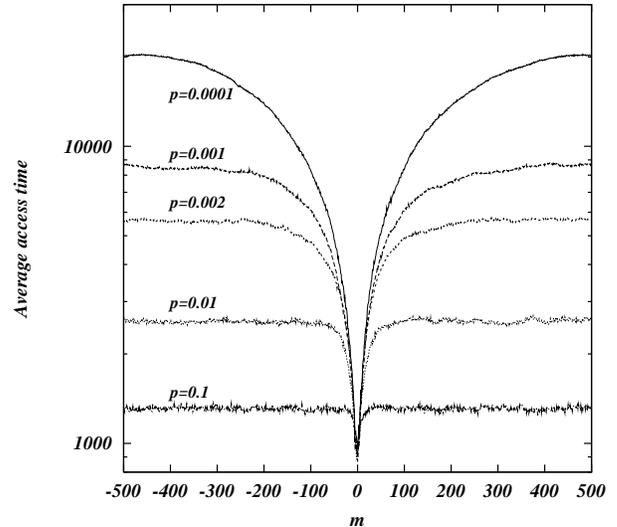}   
  \caption{ The plot of average access time vs the distance $m$ from
    the starting point $m=0$ for graph with $1000$ nodes and $k=5$, generated
    for various values of parameter $p$. The average access time
    clearly show the crossover from the regular to random behavior
    with increasing $m$. Note the logarithmic scale for the average
    access time.}
  \label{fig3}
  \end{center}
\end{figure}

Fig.~\ref{fig3} shows the results of  average access time on the
graph with $1000$ nodes and $k=5$. Two distinct behaviors of average access
time can be identified from 
the Fig.~\ref{fig3}. For small values of $m$, the behavior is similar 
to that of regular case, while for the larger values of $m$ the
average access 
time saturates and behaves like that of a random graph (See
Sec.~\ref{peqone}). As $p$ increases the saturation of average access time
becomes more prominent and the crossover from regular to saturation
behavior takes place at smaller and smaller values of $m$. For
$p=0.01$, which corresponds to the small-world
behavior~\cite{WattsStrogatz}, the average access time behavior is almost
same as that of the random graph except for small values of $m$ of the 
order of $k$. 

To obtain analytical estimates of average access time we again make use
of the independent step approximation defined in
Sec.~\ref{peqone}. In this approximation each step in the walk is taken with a
probability $(1-p)$ to nearest $2k$ sites and with probability $p$ to
the remaining sites. In analogy with the $p=0$ case we write a
recursion relation for $D_{i,m}$ (See Appendix). 

\begin{eqnarray}
  \label{eq:recurmidpk}
  D_{i,m} &=&  (1 - p) \Big[ {1 \over {2k}} \sum_{l=0}^k(D_{i+l,m} + D_{i-l,m})
  + 1 \Big] \nonumber \\
  & & + p \Big[ {1 \over {n-2k-1}}
  \!\!\!\!\!\!\!\!\!\!\!\!\!\!\sum_{\stackrel{
      l=0}{\makebox{ $l \not= {i-k,\ldots,i+k}$}}}^{n-1}
  \!\!\!\!\!\!\!\!\!\!\!\!\!\!\!D_{l,m} + 1\Big]
\end{eqnarray}

This is a $2k$-th order difference equation. As in the case of $p=0$, we
first consider the case $k=1$. The recursion relation becomes 

\begin{eqnarray}
\label{eq:recurmidp}
  D_{i,m} &=& (1 - p) \Big[ {1 \over 2} (D_{i+1,m} + D_{i-1,m}) + 1 \Big]
  \nonumber \\ & & + p
  \Big[ {1 \over {n-3}}
  \!\!\!\!\!\!\!\!\!\!\!\!\!\!\sum_{\stackrel{ l=0}{\makebox{ $i \not=
        {i,i-1,i+1}$}}}^{n-1} \!\!\!\!\!\!\!\!\!\!\!\!\!\!\!D_{l,m} + 
  1\Big]
\end{eqnarray}
which can be written as 
\begin{eqnarray}
  D_{i,m} &=& (1 - p) \Big[ {1 \over 2} (D_{i+1,m} + D_{i-1,m}) + 1 \Big]
  \nonumber \\
  & &+ p \Big[ {1 \over {n-3}} \sum_{l=0}^{n-1} D_{l,m} + 1 - {1 \over {n-3}}
  ( D_{i,m} + D_{i+1,m} + D_{i-1,m} ) \Big] \nonumber 
\end{eqnarray}
with little algebra, we finally get
\begin{eqnarray}
  \label{eq:final}
  D_{i,m} = \xi (D_{i+1,m} + D_{i-1,m}) + \zeta
\end{eqnarray}
where 
\begin{eqnarray}
  \xi = { {(n-3) - p (n-1)} \over {2 (n-3+p)} } \nonumber
\end{eqnarray}
and
\begin{eqnarray}
  \zeta = { {p \sum_{j=0}^{n-1} D_{i,m} + n - 3} \over {n-3+p}} \nonumber
\end{eqnarray}

Note that the sum $\sum_{j=0}^{n-1} D_{i,m}$ is independent of the site index
and can be treated as a constant to be determined self-consistently
from the solution.
We solve the different equation~(\ref{eq:final}) using the standard
methods~\cite{Feller}.
\begin{eqnarray}
  D_{i,m} = A \theta_+^{(m-i)} + B \theta_-^{(m-i)} - {\zeta \over
    {2\xi -1}} \nonumber
\end{eqnarray}
where
\begin{eqnarray}
  \theta_{\pm} = {1 \over {2\xi}} (1 \pm \sqrt{1 - 4 \xi^2}) \nonumber
\end{eqnarray}

The constants $A$ and $B$ are determined using the boundary conditions
$D_{m,m}=0$ and $D_{m-n,m} = 0$. The solution is given by 
\begin{eqnarray}
  D_m = {\zeta \over {2 \xi - 1}} \left[ { {\theta_-^{-m}
        \theta_+^{-n} - \theta_+^{-m} \theta_-^{-n} - \theta_-^{-m} +
        \theta_+^{-m} } \over {\theta_+^{-n} - \theta_-^{-n}} } - 1 \right]
  \label{finalDm}
\end{eqnarray}
where without loss of generality we have put the starting point as $i=0$.

The sum $\sum_{m=0}^{n-1} D_m$ occurring in $\zeta$ is calculated  by
summing Eq.~(\ref{finalDm}) for all values of $m$ and is given by
\begin{eqnarray}
 \sum_{m=0}^{n-1} D_m  = { {(n-3)
  } \over { p \left( 1 + n/ 
       \sum_{m=1}^{n-1} \left[ { {\theta_-^{-m}
        \theta_+^{-n} - \theta_+^{-m} \theta_-^{-n} - \theta_-^{-m} +
        \theta_+^{-m} } \over {\theta_+^{-n} - \theta_-^{-n}} } - 1 \right]
     \right) } } \nonumber
\end{eqnarray}

Note that Eq.~(\ref{eq:recurmidpk}) and further analysis
exhibit random behavior as $p \rightarrow {{(n-2k-1)} \over {(n-1)}}$,
rather than $p=1$ (See Appendix).    

For general value of $k$, we again write a scaling relation similar to 
that of the case $p=0$.
\begin{eqnarray}
  \label{eq:block1}
  D_m^{k}(n,p) \approx (1 + \mu(p,k) \ln(k))  D_{{m \over {k}}}^1({n
    \over k},p) 
\end{eqnarray}

\begin{figure}[hbt]
  \begin{center}
 \def \epsfsize#1#2{0.45#1}
 \figdraw{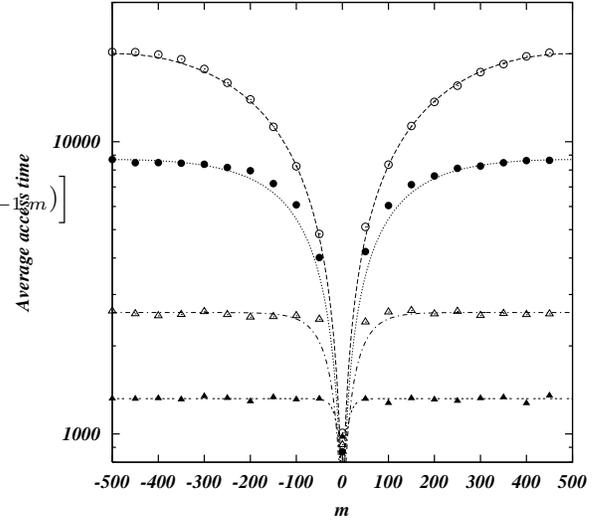}   
  \caption{ Plot of average access time vs the distance $m$ from the
    starting point of the random walk. The points are the numerical
    results for $n=1000$, $k=5$ and $p= 0.0001$ (open circle),
    $p=0.001$ (solid circle), $p=0.01$ (open triangle) and $p=0.1$
    (solid triangle). The lines are obtained using the scaling
    relation Eq.~(\ref{eq:block1}).}
  \label{fig4}
  \end{center}
\end{figure}

The Fig.~\ref{fig4} shows the match of the numerical data with the
analytic expressions. The parameter $\mu$ has a weak dependence on $k$ 
for non-zero $p$
which we have neglected in further analysis. 
\begin{figure}[hbt]
  \begin{center}
 \def \epsfsize#1#2{0.45#1}
 \figdraw{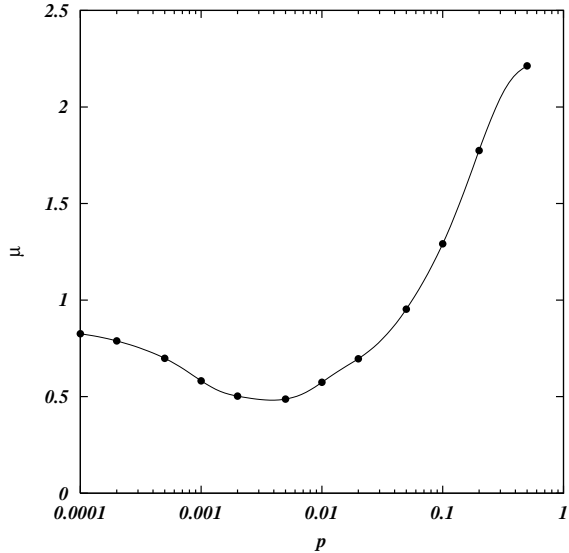}   
  \caption{ Plot of $\mu(p)$ as a function of $p$.}
  \label{fig5}
  \end{center}
\end{figure}
 Fig.~\ref{fig5} shows the behavior of $\mu$
for the various values of $p$. In Fig.~\ref{fig4} it is clearly seen that
for small and large values of $p$ the match numerical and equation is
quite good but for 
intermediate values there is a considerable deviation from the
numerical results for small values of $m$. This fact can be understood
as follows: 

For small values of $p$, i.e., when $p \approx 10^{-4}$, the graph
is nearly regular and the blocks of size $k$ are nearly completely
connected graphs, but as $p$ increases the probability of leaving the
block randomly to far away point increases, giving rise to higher
average access time for the nearby points in the block, than the
analytical values. Again for high values of $p$ the expression
has a good match because as far as the average access time is concern the
completely connected block and random block behave in similar way.

\section{Discussion}
\label{discussion}

It is interesting to compare the limiting behaviors of intermediate
cases with those of results for the regular and random cases. The
Eq.~(\ref{eq:recurmidp}) reduces to Eq.~(\ref{eq:reckeqone}) as $p
\rightarrow 0$. However, the solution (Eq.~(\ref{finalDm})) does not 
smoothly reduce to Eq.~(\ref{eq:solkeqone}). This is because of the
degeneracy in the roots of quadratic indicial equation of
Eq.~(\ref{eq:reckeqone}) for $p=0$ which is lifted for non-zero
$p$. Thus the solution changes from a polynomial to an exponential in
site index as $p$ becomes non-zero. This explains the reason behind
the saturation of the solution for non-zero $p$ for asymptotic $m$
which can not be obtained for $p=0$.

For the random case we again find a similar situation. As $p
\rightarrow { {(n-3)} \over {(n-1)}}$, which corresponds to the random
case (see Appendix), the Eq.~(\ref{eq:recurmidp}) reduces to the $D_m = \zeta$ and is
consistent with the solution $D_0 = 0$ and $D_m = n-1$, $m \not=
0$. However, the solution (Eq.~(\ref{finalDm})) does not smoothly
reduce to a constant function because one of the roots ($\theta_+$)
diverges. 

Next, we consider the average access time for the
diametrically  opposite node i.e., $G(p) = D_{n \over 2}(p)$. This
quantity is of interest as
it should be correlated to the average cover time for the graph.

\begin{figure}[hbt]
  \begin{center}
 \def \epsfsize#1#2{0.45#1}
 \figdraw{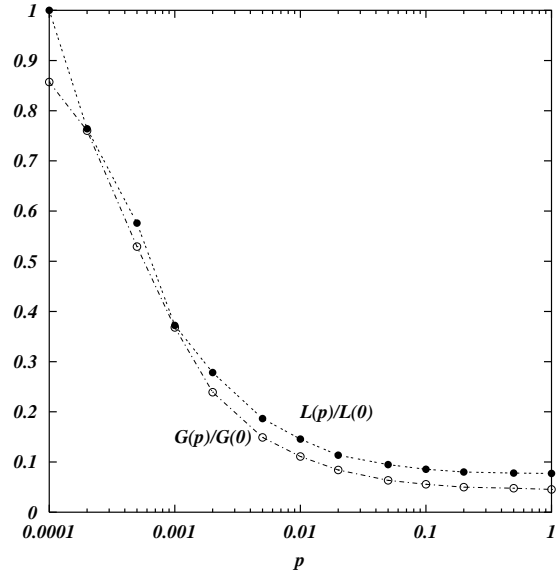}   
  \caption{ Plot of $G(p)/G(0)$, i.e., normalized average access time
    for the diametrically opposite point of the graph, vs $p$. The
    plot also shows characteristic length $L(p)/L(0)$ as a function of 
    $p$. Both the quantities show a similar behavior. In particular
    the sharp drop  near  small-world regime ($p \approx 0.01$) to the
    value corresponding to the random case is clearly seen in the
    figure.} 
  \label{fig6}
  \end{center}
\end{figure}

Fig.~\ref{fig6} shows the behavior of the access time of the
diametrically opposite point normalized with that of $p=0$ case, i.e.
$G(p)/G(0)$, as a function of $p$. The figure also shows the
graph of characteristic length $L(p)/L(0)$~\cite{WattsStrogatz}, which 
is the average shortest path between any two sites. Both the curves in
the graph corresponding to the shortest path and the random walk show
similar behavior. This observation has interesting consequences. The
random spread on a small-world network considerably reduces the
access time compared to the regular graph as in the case of the
shortest path spread. This can be very 
useful in applications such as routing and switching where random
routing is cheaper. The determination of shortest path is generally
very expensive (the number of operations goes as $n^3$) and also in
many cases the
complete information of the network is not available. Examples of such 
networks are social networks, telephone networks, internet etc. where
the number of nodes is very large. In such cases random routing will
be more effective and cheaper, particularly if the graph has the
small-world geometry. In this case the average access time goes as
$O(n)$. 

In the calculations we have introduced an {\em independent step
  approximation}, where we average over the various realizations of
graphs at each time step. This is a kind of mean field approximation
and has allowed us to obtain recursion relations for the average
access time. We expect this approximation to be reasonably good for
the random case. Even for the intermediate cases, the results obtained
by this approximation are in good agreement with the numerical values. 

For general values of $k$ the recursion relations for average access
time can not be solved. However, the behavior of average access time
shows an interesting scaling relation in $k$. As discussed in
Sec.~\ref{peqzero} the scaling relation corresponds to a quotienting
procedure where a graph of $n$ nodes with $2k$ connections is scaled
to a graph of $ {m \over k}$ nodes with nearest neighbor
connections and preserving the far-edges~\cite{Sag}. The scaling gives 
very good fit to the numerical data except for small values of $m$,
when $p \geq 0.001$.

\section{Summary}
\label{summary}

In this paper, we have studied random walk on the family of
small-world graphs. For the regular case the average access time shows 
linear behavior for small distances and shows quadratic nature for
large distances. An interesting scaling relation in $k$ is observed
for the average access time. For the random case the average access
time is $(n-1)$ and is independent of distance and $k$. An independent
step approximation has enabled us to get the analytical result for the
average access time. The same approximation allows us to write a
recursion relation for the intermediate values of $p$. For
intermediate cases the average access time shows a crossover from
regular to random behavior with increasing distances. 

The normalized average access time of the diametrically opposite nodes
shows almost identical behavior as that of the characteristic length
as a function $p$. This observation can be very important in several
applications where the number of nodes in graph are very large or the
complete information about the graph is not available. In these cases
random routing or switching will be beneficial and cheaper.

We thank Prof. V. Balakrishnan, Prof. G. Baskaran, Dr. G. Menon and
Dr. R. Ramanujam for fruitful discussions. 

\section {Appendix}
\label{appendix}
\subsection {Derivation of equation~(\ref{eq:recurmidp})}

We assume that the probability of breaking of an edge is $p$.
We identify that the contribution to the average access time of site
$j$ comes from three types of events.
\begin{enumerate}
\item All the $2k$ neighbors of site $j$ are connected to $j$. The
  probability associated with this event is $(1-p)^{2k}$.
\item Some of the neighbors are connected to $j$ (say $2k - r$) and
  $r$ are connected to the far away sites. The probability associated
  with this event is $p^r (1-p)^{2k-r}$. 
\item None of the neighbors of $j$ are connected to $j$. The
  probability associated with this event is $p^{2k}$.
\end{enumerate}

By independent step approximation the degree of each node is
$2k$. This enables up to write a recursion relation for average access 
time using the properties of expectation value of conditional
probability~\cite{Feller}. We write 
\begin{eqnarray}
  \label{eq:step1}
  D_{i,m} &=& \sum_{r=0}^{2k} \makebox{ }^{2k}C_r p^r (1-p)^{2k-r} \left\{ { {2k
        - r - 1 } \over {2k} } \left[ \sum_{l=1}^{k} \left(
        D_{i-l,m} + D_{i+l,m} \right) \right] + {r \over {n - 2k - 1}} 
\!\!\!\!\!\!\!\!\!\!\!\!\!\!\sum_{\stackrel{
      l=0}{\makebox{ $l \not= {i-k,\ldots,i+k}$}}}^{n-1}
  \!\!\!\!\!\!\!\!\!\!\!\!\!\!\!D_{l,m} \right\} + 1\\
&=& (1 - p) \Big[ {1 \over {2k}} \sum_{l=0}^k(D_{i+l,m} + D_{i-l,m})
  + 1 \Big] + p \Big[ {1 \over {n-2k-1}}
  \!\!\!\!\!\!\!\!\!\!\!\!\!\!\sum_{\stackrel{
      l=0}{\makebox{ $l \not= {i-k,\ldots,i+k}$}}}^{n-1}
  \!\!\!\!\!\!\!\!\!\!\!\!\!\!\!D_{l,m} + 1\Big]
\end{eqnarray}

In the above calculation we have assumed that the edge broken with
probability $p$ is not rewired to one of the $2k$ nearest
neighbors. Due to this the random graph limit corresponds to $p=
{{(n-2k-1)} \over {(n-1)}}$ instead of $p=1$. This can be seen by
equating the probability of a connection to a nearest neighbor to the
probability of a connecting to any of the other sites
\begin{eqnarray}
  { {(1-p)} \over {2k} } = { p \over {n-2k-1} }. \nonumber
\end{eqnarray}

\end{document}

%% file: figures/figdef.tex
\def \figdraw #1 {        \newbox\boxtmp
        \setbox\boxtmp=\hbox{\epsfbox{#1}}
        \usebox{\boxtmp} \\}
\def \putname #1 {\vfil \rightline{#1}}

%% file: preprint.bbl
\begin{thebibliography}{99}
\bibitem{WattsStrogatz} {D. J. Watts, S. H. Strogatz, Nature
{\bf 393} , 440 (1998).}
\bibitem{Sag} {S. A. Pandit,
  R. E. Amritkar, Phys. Rev. E {\bf 60} (2), R1119 (1999).}
\bibitem{NewmanWatts}{M. E. J. Newman, D. J. Watts, Scaling and
  percolation in the small-world network model, preprint {\tt
    cond-mat/9904419}.} 
\bibitem{Moukarzel} {C. F. Moukarzel, Preading and shortest paths in
    systems with sparse long-range connections, preprint {\tt
      cond-mat/9905322}.}  
\bibitem{MooreNewman} {C. Moore, M. E. J. Newman, Epidemics and
    percolation in small-world networks, preprint.}
\bibitem{Milgram}{S. Milgram, Psychol. Today {\bf 2}, 60-67 (1967).} 
\bibitem{Stevens}{W. R. Stevens, {\it Unix network programing},
    (Prantice Hall, India (1993))}
\bibitem{Aho}{A. V. Aho, J. E. Hopcroft, J. D. Ullman, {\it The design 
      and analysis of computer algorithms}, (Addison-Wesley (1998)).}
\bibitem{Lovasz}{ L. Lov\'asz, Combinatorics, Paul Erod\"os is Eighty
    (vol. 2), 1-46 (1993).}
\bibitem{Feller} {W. Feller, {\it An introduction to probability
      theory and its applications, Vol. 1}, (Wiley Eastern limited,
    Ed. 3, (1968)).}
\end{thebibliography}
